\documentclass[conference,twocolumn]{IEEEtran}

\usepackage{amsmath}
\usepackage{amsthm}
\usepackage{amssymb}
\usepackage{graphicx}
\usepackage{subfig}
\usepackage{rotating}
\usepackage{flushend}
\usepackage{verbatim}
\allowdisplaybreaks

\newsavebox{\tempbox}

\usepackage{xcolor}

\newcommand{\logt}{\log_{2}}

\newcommand{\abs}[1]{\left| #1 \right| }

\newcommand{\labs}[1]{\|{#1}\|_{1}}
\newcommand{\defn}{\triangleq}

\newcommand{\mbf}[1]{\mathbf{#1}}
\newcommand{\mcf}[1]{\mathcal{#1}}

\newcommand{\cexp}[1]{\frac{1}{n} \logt {#1} }

\newcommand{\set}[1]{\left\{ #1 \right\}}
\newcommand{\eo}[1]{\doteq_{\tiny{#1}}}

\DeclareMathOperator{\markov}{\setlength{\unitlength}{.5cm} \begin{picture}(1,1)  \put(0,.22){\line(1,0){1}}  \put(.5,.22){\circle{.3}}   \end{picture}}

\newtheorem{theorem}{Theorem}
\newtheorem*{fano}{Strong Fano's inequality}
\newtheorem*{stable}{Information stabilization}

\newtheorem{lemma}[theorem]{Lemma}

\IEEEoverridecommandlockouts

\begin{document}
\title{Wiretap channel capacity: Secrecy criteria, strong converse,
  and phase change}
\author{Eric Graves and Tan F. Wong
\thanks{Eric Graves is with Army Research Lab,  Adelphi, MD
  20783, U.S.A. \texttt{ericsgra@ufl.edu} }%
\thanks{Tan F. Wong is with Department of Electrical and Computer Engineering,
 University of Florida,
 Gainesville, FL 32611, U.S.A. 
\texttt{twong@ufl.edu}}%
\thanks{T. F. Wong was supported by the National Science Foundation
under Grant CCF-1320086. Eric Graves was supported by a National
Research Council Research Associateship Award at Army Research Lab.}
}

\maketitle

\begin{abstract}
  This paper employs equal-image-size source partitioning techniques
  to derive the capacities of the general discrete memoryless wiretap
  channel (DM-WTC) under four different secrecy criteria. These
  criteria respectively specify requirements on the expected values
  and tail probabilities of the differences, in absolute value and in
  exponent, between the joint probability of the secret message and
  the eavesdropper's observation and the corresponding probability if
  they were independent.  Some of these criteria reduce back to the
  standard leakage and variation distance constraints that have been
  previously considered in the literature. The capacities under these
  secrecy criteria are found to be different when non-vanishing error
  and secrecy tolerances are allowed. Based on these new results, we
  are able to conclude that the strong converse property generally
  holds for the DM-WTC only under the two secrecy criteria based on
  constraining the tail probabilities. Under the secrecy
  criteria based on the expected values, an interesting phase change
  phenomenon is observed as the tolerance values vary.
\end{abstract}

\section{Introduction}
The discrete memoryless wiretap channel (DM-WTC)
$(\mcf{X}, P_{Y,Z|X},\mcf{Y}\times \mcf{Z})$ consists of a sender $X$,
a legitimate receiver $Y$, and an eavesdropper $Z$. A message $M$ is
to be sent reliably from $X$ to $Y$ and discreetly against
eavesdropping by $Z$.  Over $n$ uses of the DM-WTC, let
$f^n : \mcf{M} \rightarrow \mcf{X}^n$ and
$\varphi^n: \mcf{Y}^n \rightarrow \mcf{M}$ be the encoding and
decoding functions respectively employed at $X$ and $Y$, where
$\mcf{M} = [1:2^{nR}]$ is the message set and $M$ is uniformly
distributed over $\mcf{M}$. The transmission 
reliability requirement is specified by
\begin{equation} \label{eq:pe}
\Pr \left\{ \varphi^n( Y^n ) \neq M  \right\} \leq \epsilon_n
\end{equation}
where $\epsilon_n \in (0,1)$ denotes the error tolerance.  The secrecy
requirement 
assesses how much one may learn about $M$ from $Z^n$.  This
requirement is often quantified by measuring the level of
``independence'' between $M$ and $Z^n$ based on either the variation
distance
\begin{align*}
&\hspace*{-10pt}
\labs{P_{M,Z^n} - P_{M}P_{Z^n}} 
\\
&\defn 
\frac{1}{2}\sum_{(m,z^n) \in \mcf{M} \times
  \mcf{Z}^n} \abs{P_{M,Z^n}(m,z^n) - P_{M}(m)P_{Z^n}(z^n)}
\end{align*}
or the divergence $D(P_{M,Z^n} \| P_{M}P_{Z^n}) = I(M;Z^n)$ between
$P_{M,Z^n}$ and $P_{M}P_{Z^n}$. Another way of quantifying the secrecy
requirement is to view the problem as a binary hypothesis testing of
the alternate hypothesis of $M$ and $Z^n$ being independent against
the null hypothesis of $M$ and $Z^n$ being correlated. This is an
interesting case in which we would like the false positive probability
given by the likelihood ratio test
\begin{align*}
&\hspace*{-10pt}
P_{M,Z^n} \left( \left\{ (m,z^n) \in \mcf{M}\times\mcf{Z}^n: 
    \frac{P_{M}(m)P_{Z^n}(z^n)}{P_{M,Z^n} (m,z^n) }\geq \tau \right\}
  \right)
\\
  & \rightarrow 1 \text{~~~as } n \rightarrow \infty
\end{align*}
    \footnote{Hereafter, convergence of any quantity indexed by $n$
    means convergence as $n \rightarrow \infty$. For example,
    $\delta_n \rightarrow 0$ means $\delta_n$ converges to $0$ as
    $n \rightarrow \infty$.}where the decision threshold $\tau \in [0,1)$ serves as a measure of
secrecy with $\tau \rightarrow 1$ being the most secret situation.
Note that the log-likelihood
$\logt \frac{P_{M}(m)P_{Z^n}(z^n)}{P_{M,Z^n} (m,z^n)}$ may also be used
  in the hypothesis testing problem above.

For every $(m,z^n) \in \mcf{M}\times\mcf{Z}^n$, define
\[
{v} (m,z^n) \defn \begin{cases}
  \left[1-\frac {P_{M}(m)P_{Z^n}(z^n)}{P_{M,Z^n} (m,z^n)} \right]^{+}  &
  \text{if } P_{M,Z^n}(m,z^n) >0 \\
  0 & \text{if } P_{M,Z^n}(m,z^n) =0
\end{cases}
\]
where $[c]^+$ equals $c$ if $ c> 0$ and $0$ otherwise, and 
\[
{i} (m,z^n) \defn 
\begin{cases}
  -\logt \frac{P_M(m)P_{Z^n}(z^n)}{ P_{M,Z^n} (m,z^n)}
  &
  \text{if } P_{M,Z^n}(m,z^n) >0 \\
  0 & \text{if } P_{M,Z^n}(m,z^n) =0.
\end{cases}
\]
All the secrecy requirements discussed above can be
compactly specified in terms of the tail probabilities and expected
values of $v(M,Z^n)$ and $i(M,Z^n)$: 
\begin{align*}
\mbf{S}_1(\delta_n) &:P_{M,Z^n} \left( \set{
                      {v}(M,Z^n) > \delta_n }\right) \leq \mu_n  
\text{ for some } \mu_n \rightarrow 0
\\
\mbf{S}_2(\delta_n) &:E_{M,Z^n} \left[ v(M, Z^n) \right]
= \labs{P_{M,Z^n} - P_{M}P_{Z^n}} 
\leq \delta_n
\\
\mbf{S}_3(l_n) &: P_{M,Z^n} \left( \set{
                   {i} (M,Z^n) > l_n }\right) \leq \mu_n
\text{ for some } \mu_n \rightarrow 0
\\
\mbf{S}_4(l_n) &: E_{M,Z^n}\left[  {i}(M, Z^n) \right] 
= I(M;Z^n) \leq l_n
\end{align*}
where $\delta_n \in (0,1]$, $l_n\in (0, \infty)$, and
$E_{M,Z^n}[\cdot]$ denotes the expectation w.r.t. $P_{M,Z^n}$.  Note
that $\mbf{S}_2(\delta_n)$ and $\mbf{S}_4(l_n)$ are the variation
distance and divergence (leakage) constraints, respectively, while
$\mbf{S}_1(\delta_n)$ and $\mbf{S}_3(l_n)$ correspond to the secrecy
requirements specified by the hypothesis testing problem using the
likelihood and log-likelihood ratios, respectively.

Clearly these secrecy requirements are related to each other. For
example, we have
$\mbf{S}_1(\delta_n) = \mbf{S}_3\left( -\logt
  (1-\delta_n)\right)$. Also, $\mbf{S}_1(\delta_n)$ implies
$\mbf{S}_2(\delta_n+\mu_n)$.  By Markov's inequality,
$\mbf{S}_2(\delta_n)$ implies $\mbf{S}_1(\sqrt{\delta_n})$ if
$\delta_n \rightarrow 0$.  Thus for vanishing tolerances (i.e.,
$\delta_n \rightarrow 0$), $\mbf{S}_1$, $\mbf{S}_2$, and $\mbf{S}_3$
are essentially equivalent. In addition, by Pinsker's inequality,
$\mbf{S}_4(l_n)$ implies
$\mbf{S}_2\left(\sqrt{\frac{l_n \ln 2}{2}} \right)$ if
$l_n \in (0,\frac{2}{\ln 2})$.

Special cases of these secrecy requirements have been considered in
the literature. For example, requiring $\epsilon_n \rightarrow 0$
in~\eqref{eq:pe}, $\mbf{S}_4(nr_l)$ is the equivocation constraint
originally considered in~\cite{Wyner1975}. Six secrecy requirements
$\mathbb{S}_1$--$\mathbb{S}_6$ are more recently
considered\footnote{Note that $\mathbb{S}_5$ in~\cite{bloch13strong}
  seems problematic as it can always be trivially satisfied.}
in~\cite{bloch13strong}. Setting $\epsilon_n \rightarrow 0$,
$\mathbb{S}_1$ is $\mbf{S}_4(l_n)$ for some $l_n \rightarrow 0$,
$\mathbb{S}_2$ is $\mbf{S}_2(\delta_n)$ for some
$\delta_n \rightarrow 0$, $\mathbb{S}_3$ is $\mbf{S}_3(l_n)$ for some
$l_n \rightarrow 0$, $\mathbb{S}_4$ is $\mbf{S}_4(l_n)$ for some
$\frac{l_n}{n} \rightarrow 0$, and $\mathbb{S}_6$ is $\mbf{S}_3(l_n)$
for some $\frac{l_n}{n} \rightarrow 0$.

The majority of known secrecy capacity results under the above secrecy
requirements are for cases with vanishing error tolerance,
$\epsilon_n \rightarrow 0$, and secrecy tolerance,
$l_n \rightarrow 0$, $\frac{l_n}{n} \rightarrow 0$, or
$\delta_n \rightarrow 0$. These results are nicely summarized
in~\cite{bloch13strong}, which shows that the secrecy capacities under
$\mathbb{S}_1$--$\mathbb{S}_6$ (see footnote~2) of the DM-WTC are all
given by $\max_{P_{U,X }} I(U;Y) - I(U;Z)$, where
$U \markov X \markov Y,Z$.  Here we are mainly interested in cases
where both the error tolerance $\epsilon_n$ and secrecy tolerance
$\delta_n$, $l_n$ or $\frac{l_n}{n}$ are non-vanishing, on which only
a few partial results exist.  The oldest such result dates back to
Wyner's original paper~\cite{Wyner1975}, in which the secrecy capacity
under $\mbf{S}_4(nr_l)$, where $r_l>0$ denotes the leakage rate, of
the degraded DM-WTC ($P_{Y,Z|X} = P_{Z|Y}P_{Y|X}$) is calculated for
the case of $\epsilon_n \rightarrow 0$.  The $\epsilon$-secrecy
capacity under $\mbf{S}_4(l_n)$ of the degraded DM-WTC is obtained
in~\cite{tan14wiretap} for the case of $\frac{l_n}{n} \rightarrow 0$.
This case has also been extended to the general DM-WTC
in~\cite{graves2015equal} and~\cite{2016arXiv161004215W}. The
$\epsilon$-secrecy capacity under $\mbf{S}_2(\delta)$ of the degraded
DM-WTC is found in~\cite{Hayashi14}.

In this paper, we determine the secrecy capacities for the general
DM-WTC under the above four security requirements,
$\mbf{S}_1$--$\mbf{S}_4$, with non-vanishing tolerances. The converses
of all of these capacity results are new, and are straightforwardly
obtained using our recently developed equal-image-size source
partitioning
techniques~\cite{graves2015equal,graves2016information}. Further, the
$\epsilon$-secrecy capacity for each of these four requirements is
unique. Under $\mbf{S}_1$ and $\mbf{S}_3$ the \emph{strong converse}
property holds, while it does not under $\mbf{S}_2$ and $\mbf{S}_4$ in
general. In addition, under $\mbf{S}_2$ and $\mbf{S}_4$, the capacity
can be broken into distinct phases depending on the error
tolerance. For instance, under $\mbf{S}_2$ the capacity of the channel
is either equal to the capacity of the channel with vanishing error,
or the capacity of the channel with no secrecy requirement. We call
this interesting phenomenon a \emph{phase change.}

\section{Main results} \label{sec:main} 

For $i \in \set{1,2,3,4}$, we call
  $(f^n,\varphi^n)$ a $(n,R_n,\epsilon_n, \mbf{S}_i(\eta_n))$-code if
  the domain of $f^n$ (i.e., $\mcf{M}$) is of cardinality $2^{nR_n}$,
  and the pair satisfy both~\eqref{eq:pe} and $\mbf{S}_i(\eta_n)$.
  Further we say the \emph{rate error secrecy} (RES)-triple
  $(a,b,c) \in \mathbb{R}^3$ is $\mbf{S}_i$-achievable if there exists
  a sequence of $(n, R_n, \epsilon_n, \mbf{S}_i(\eta_n))$-codes such
  that
  $\lim_{n \rightarrow \infty} (R_n,\epsilon_n, \eta_n) = (a,b,c)$ if
  $i \in \set{1,2}$, and
  $\lim_{n \rightarrow \infty} \left(R_n,\epsilon_n, \frac{\eta_n}{n}
  \right) = (a,b,c)$ if $i \in \set{3,4}$. Then the
  $\epsilon$-secrecy capacity under the appropriate $\mbf{S}_i(\cdot)$
  is the maximum $R$ such that the RES-triple $(R,\epsilon,\eta)$ is
  $\mbf{S}_i$-achievable. 
  
  Note that for $\mbf{S}_3$ and $\mbf{S}_4$, the above definition
  corresponds to what is called ``weak'' secrecy in the
  literature~\cite{bloch13strong}. If ``strong'' secrecy is desired,
  the definition could be modified to that the RES-triple
  $(R,\epsilon,\eta)$ is $\mbf{S}_i$-achievable when there exists a
  sequence of $(n, R_n, \epsilon_n, \mbf{S}_i(\eta_n))$-codes such
  that
  $\lim_{n \rightarrow \infty} (R_n,\epsilon_n, \eta_n) =
  (R,\epsilon,\eta)$, for $i \in \set{3,4}$.
  We have instead chosen to present the ``weak'' versions of these
  criteria, simply because their proofs trivially recover their
  ``strong'' counterparts. 

%
Write $C(r_l)$ to denote the capacity of the wiretap channel subject to the weak leakage constraint $r_l \geq n^{-1}I(Z^n;M)$. In specific,
\begin{align*}
&C(r_l) = \\
&\max_{P_{W,U,X}}  \min \left(I(Y;U|W) - I(Z;U|W) + r_l, I(Y;U,W) \right),
\end{align*}
where $\abs{\mcf{U}} \leq (\abs{\mcf{X}}+1)(\abs{\mcf{X}}+3)$ and $\abs{\mcf{W}} \leq \abs{\mcf{X}}+3$. Two values of distinction which will arise in our results are that of $C(0)$ and $C(\infty)$ for which
\begin{align*}
C(0) &= \max_{P_{U,X}} I(Y;U) - I(Z;U),\\
C(\infty) &= \max_{P_{X}} I(Y;X) .
\end{align*}
Next, restrict
$\epsilon \in [0,1)$ and $\delta \in [0,1]$, and $r_l \in [0,\infty)$.
Then the following theorems give our main results regarding the
secrecy capacities:
\begin{theorem} \label{thm:1}
The $\epsilon$-secrecy capacity under $\mbf{S}_1(\delta)$ of the
DM-WTC is given by 
\[
\mathbb{C}_{1}(\delta) \defn \begin{cases}
C(0) &\text{if } \delta < 1 \\
C(\infty) &\text{otherwise}.
\end{cases}
\]
 for all $\epsilon$.
\end{theorem}
\begin{theorem} \label{thm:2}
  The $\epsilon$-secrecy capacity under $\mbf{S}_2(\delta)$ of the
  DM-WTC is given by
\[
\mathbb{C}_{2}(\epsilon,\delta) \defn \begin{cases}
C(0) &\text{if } \epsilon + \delta < 1 \\
C(\infty) &\text{otherwise}.
\end{cases}
\]
\end{theorem}
\begin{theorem} \label{thm:3} 
  The $\epsilon$-secrecy capacity under $\mbf{S}_3(nr_l)$ of the
  DM-WTC is given by
\[
\mathbb{C}_3(r_l) \defn C(r_l) 
\]
for all $\epsilon$.
\end{theorem}
\begin{theorem} \label{thm:4} 
  The $\epsilon$-secrecy capacity under $\mbf{S}_4(nr_l)$ of the
  DM-WTC is given by
\[ 
\mathbb{C}_{4}(\epsilon,r_l) \defn C\left( \frac{r_l}{1-\epsilon} \right) .
\]
\end{theorem}
As mentioned before, the main new contributions are the converses of
the theorems. Theorem~\ref{thm:2} extends the result
in~\cite{Hayashi14} from the degraded DM-WTC to the general DM-WTC.
Theorems~\ref{thm:3} and~\ref{thm:4} extend the results
in~\cite{bloch13strong} and
in~\cite{graves2015equal,2016arXiv161004215W} to the case of
$r_l > 0$, respectively.

Theorems~\ref{thm:1} and~\ref{thm:3} state that the $\epsilon$-secrecy
capacities of the DM-WTC under $\mbf{S}_1(\delta)$ and
$\mbf{S}_3(nr_l)$ are invariant to the value of $\epsilon \in [0,1)$
for all valid values of $\delta$ and $r_l$, respectively. In other
words, the strong converse property holds under $\mbf{S}_1(\delta)$
and $\mbf{S}_3(nr_l)$. Although invariant of the error tolerance, the
$\epsilon$-secrecy capacity under $\mbf{S}_3(nr_l)$ is non-trivially
dependent on the leakage rate $r_l$.
In specific, the $\epsilon$-secrecy capacity under $\mbf{S}_3(nr_l)$
increases linearly as a function of $r_l$ from $C(0)$ until it
saturates at $C(\infty)$, the (non-secret) capacity of the discrete
memoryless channel (DMC) $(\mcf{X}, P_{Y|X}, \mcf{Y})$.

For the secrecy requirements $\mbf{S}_2(\delta)$ and
$\mbf{S}_4(nr_l)$, Theorems~\ref{thm:2} and~\ref{thm:4} respectively
show that the strong converse property no longer holds for the DM-WTC
as the $\epsilon$-secrecy capacities generally depend on the value of
$\epsilon$.
Under $\mbf{S}_2(\delta)$, the $\epsilon$-secrecy capacity remains at
$C(0)$ as long as $\epsilon \in [0, 1-\delta)$. However, for
$\epsilon \in [1-\delta,1)$, the $\epsilon$-secrecy capacity value
experiences an abrupt phase change, increasing to $C(\infty)$ as if there is
no secrecy requirement. Restricting to within either of the two value
ranges, the $\epsilon$-secrecy capacity under $\mbf{S}_2(\delta)$ is
invariant to $\epsilon$.
%

Under $\mbf{S}_4(nr_l)$, the $\epsilon$-secrecy capacity remains at
$C(0)$ when $r_l = 0$ for all $\epsilon \in [0,1)$. Note that this also
includes the cases of strong secrecy ($\mbf{S}_4(l_n)$ with
$l_n \rightarrow 0$) and bounded leakage ($\mbf{S}_4(l_n)$ with
$l_n = l$). Thus the strong converse property holds when $r_l=0$ as
proven in~\cite{graves2015equal} and~\cite{2016arXiv161004215W}. For
any fixed $r_l \in (0,C(\infty) -C(0) )$, the $\epsilon$-secrecy capacity
increases from $C(r_l)$ to $C(\infty)$ and then levels off
as $\epsilon$ increases in the range $[0,1)$. The DM-WTC exhibits a
phase change from where the strong converse property holds to where it
does not. When $r_l \geq C(\infty)-C(0)$, the $\epsilon$-secrecy capacity
value remains at $C(\infty)$ for all $\epsilon \in [0,1)$, and the DM-WTC
exhibits another phase change after which the strong converse property
holds again.

\section{Proofs of Theorems}
\label{sec:proofs}

We prove the converses in Theorems~\ref{thm:1}--\ref{thm:4} by
employing the following strong Fano's inequality developed
in~\cite{graves2015equal} and information stabilization result
developed in~\cite{graves2016information}:
\begin{fano}
  For any $(f^n,\varphi^n)$ of rate $R$ that gives
  $\Pr\{\varphi^n(Y^n) \neq M\} \leq \epsilon$ over the DM-WTC, there
  exist a random index $Q_n$ (correlated with $M$, $Y^n$, and $Z^n$)
  that ranges over an index set $\mcf{Q}_n$ whose cardinality is at
  most polynomial in $n$, $\zeta_n \rightarrow 0$, and an index subset
\[
\mcf{Q}^R_n \defn 
\set{ q_n \in \mcf{Q}_n : R \leq \frac{1}{n} I(M;Y^n|Q_n=q_n) + \zeta_n }
\]
satisfying $P_{Q_n}( \mcf{Q}^R_n ) \geq 1 - \epsilon-\zeta_n$.
\end{fano}
\begin{stable}
  For the $(f^n,\varphi^n)$ pair, random index $Q_n$, and index set
  $\mcf{Q}_n$ above, there exist $\xi_n \rightarrow 0$ and an index
  subset $\mcf{Q}_n^Z \subseteq \mcf{Q}_n$ satisfying $P_{Q_n}(\mcf{Q}^Z_n) \geq 1 - \xi_n$:\footnote{For
    any non-negative $\lambda_n \rightarrow 0$, $a_n>0$, and $b_n>0$,
    $a_n \eo{\lambda_n} b_n$ means
    $\abs{\cexp{a_n} - \cexp{b_n}} \leq \lambda_n$.}
\begin{enumerate}
\item \label{eq:c2}
  $P_{Z^n|Q_n}( \mcf{\hat Z}^n(q_n)|q_n) \geq 1-\xi_n$, where
  $\mcf{\hat Z}^n(q_n) \defn \big\{ z^n \in \mcf{Z}^n :
  P_{Z^n|Q_n}(z^n|q_n) \eo{\xi_n} 2^{-H(Z^n|Q_n=q_n)} \big\}$,
\item \label{eq:c3} there exists a
  $\mcf{\tilde M}(q_n) \subseteq \mcf{M}$ satisfying
  $P_{M|Q_n}(\mcf{\tilde M}(q_n) |q_n) \geq 1 - \xi_n$,
  and
  $P_{M|Q_n}(m|q_n) \eo{\xi_n} 2^{-H(M|Q_n=q_n)}$ for each
  $m \in \mcf{\tilde M}(q_n)$, and
\item \label{eq:c4}
  $P_{Z^n|M,Q_n}( \mcf{\tilde Z}^n(m,q_n)|m,q_n) \geq 1-\xi_n$
  where
  $\mcf{\tilde Z}^n(m,q_n) \defn \big\{ z^n \in \mcf{Z}^n :
  P_{Z^n|M,Q_n}(z^n|m,q_n) \eo{\xi_n} 2^{-H(Z^n|M,Q_n=q_n)}
  \big\}$,
\end{enumerate}
for each $q_n \in \mcf{Q}_n^Z$.
\end{stable}
Obtained through the information stabilization result in the appendix,
the following lemma will also be needed:
\begin{lemma} \label{lem:apx} 
  For any $r \geq 0$, there exist $\tau_n \rightarrow 0$,
  $\mu_n \rightarrow 0$, and $\lambda_n \rightarrow 0$ satisfying
  $n \lambda_n \rightarrow \infty$ such that by defining
\[
  \mcf{Q}_n^S(r) \defn \set{ q_n \in \mcf{Q}_n: \frac{1}{n}
    I(M;Z^n|Q_n=q_n) \leq r+\tau_n }
\]
and
\begin{align*}
  \Omega_n(r)
  &\defn 
    \Big\{ (m,z^n) \in \mcf{M} \times \mcf{Z}^n:  
    \notag \\
  & \hspace*{40pt}
    P_{M,Z^n}(m,z^n)
   \leq  2^{n (r+\lambda_n)} P_{M}(m) P_{Z^n}(z^n) \Big\} ,
\end{align*}
then
\begin{equation*}
  P_{M,Z^n} \left( \Omega_n(r) \right) \leq 
  P_{Q_n}\left( \mcf{Q}_n^S(r) \right) + \mu_n.
\end{equation*}
\end{lemma}

For proving achievability in Theorems~\ref{thm:2}
  and~\ref{thm:4}, we will make use of the following lemma to simplify
  discussions:
\begin{lemma} \label{lem:only_other_lemma} 
  For $i \in \set{2,4}$, if the RES-triple $(R,0,\eta)$ is
  $\mbf{S}_i$-achievable, then the RES-triple
  $(R, \gamma , (1-\gamma) \eta )$ is also $\mbf{S}_i$-achievable for
  any $\gamma \in [0,1)$.
\end{lemma}


\subsection{Proof of Theorem~\ref{thm:1}}

\textbf{(Direct) }For any $\delta \in [0,1)$ and $\epsilon \in [0,1)$,
the RES-triple $(C(0),\epsilon,\delta)$ being $\mbf{S}_1$-achievable
follows directly from~\cite[Theorem~17.11]{CK}, which in particular
shows the RES-triple $(C(0),0,0)$ is $\mbf{S}_1$-achievable. On the
other hand, the RES-triple $(C(\infty),0,1)$ is $\mbf{S}_1$-achievable since
$C(\infty)$ is the channel capacity for the DMC $(\mcf{X},P_{Y|X},\mcf{Y})$,
and $\delta = 1$ corresponds to no secrecy constraint.

\textbf{(Converse)} To prove that $\mbf{C}_1(\delta)$ is an upper
bound on the $\epsilon$-secrecy capacity under $\mbf{S}_1(\delta)$,
first apply Lemma~\ref{lem:apx} to obtain values $\tau_n$, $\mu_n$,
and $\lambda_n$ which converge to $0$ as $n$ increases, such that
$P_{M,Z^n}\left( \Omega_n(0) \right) \leq P_{Q_n}\left( \mcf{Q}_n^S(0)
\right) + \mu_n$, for sets $\Omega_n(0)$ and $\mcf{Q}_n^S(0)$ as
defined in Lemma~\ref{lem:apx}. We also have that
$P_{M,Z^n}\left( \Omega_n(0) \right) \geq 1 - \rho_n $ for some
$\rho_n \rightarrow 0$,
due to $\mbf{S}_1(\delta)$.
Thus $\mbf{S}_1(\delta)$ and
Lemma~\ref{lem:apx} together imply that
\begin{equation}
P_{Q_n}\left( \mcf{Q}_n^S(0) \right) \geq P_{M,Z^n}\left(\Omega_n(0)
\right) -\mu_n \geq 1 - \rho_n -\mu_n.
\label{eq:sc1}
\end{equation}
But then the strong Fano's inequality and~\eqref{eq:sc1} together give
the existence of a $q_n \in \mcf{Q}_n$ such that
\begin{align}
R &\leq\frac{1}{n} I(M;Y^n|Q_n=q_n) + \zeta_n  \label{eq:wtc_1}\\
\frac{1}{n} I(M;Z^n|Q_n=q_n)  &\leq  \tau_n \label{eq:wtc_2}
\end{align}
since
$P_{Q_n}\left( \mcf{Q}_n^R \cap \mcf{Q}_n^S(0)\right) \geq 1 -
\epsilon - \zeta_n - \rho_n - \mu_n> 0$ for large enough $n$ and
$\epsilon \in [0, 1)$. Combining Equations~\eqref{eq:wtc_1} and~\eqref{eq:wtc_2} gives 
\[
R \leq C(0) + \zeta_n + \tau_n
\] 
for all $\epsilon \in [0,1)$.
On the other hand,
when $\delta=1$, the strong Fano's inequality (i.e.,~\eqref{eq:wtc_1})
gives 
\[
R \leq C(\infty) + \zeta_n
\] 
for all $\epsilon \in [0,1)$, as in the standard
strong converse argument for the DMC $(\mcf{X}, P_{Y|X}, \mcf{Y})$.

\subsection{Proof of Theorem~\ref{thm:2}}


\textbf{(Direct)} The RES-triple $(C(0),\epsilon,\delta)$ is $\mbf{S}_2$-achievable, once again, by~\cite[Theorem~17.11]{CK}, for $\epsilon + \delta < 1$. For $\epsilon+\delta\geq 1$, the RES-triple $(C(\infty),\epsilon,1-\epsilon)$ is $\mbf{S}_2$-achievable by Lemma~\ref{lem:only_other_lemma}, since the RES-triple $(C(\infty),0,1)$ is $\mbf{S}_2$-achievable. 

\textbf{(Converse)}
On the other hand, to prove that $\mathbb{C}_{2}(\epsilon, \delta)$ is an upper
bound on the $\epsilon$-secrecy capacity under $\mbf{S}_2(\delta)$, 
observe that $\mbf{S}_2(\delta)$ implies
\begin{align}
\delta 
&\geq 
\labs{P_{M,Z^n} - P_{M} P_{Z^n}}
\notag \\
&\geq
\sum_{(m,z^n) \in \mcf{M} \times \mcf{Z}^n \setminus \Omega_n(0) } 
  P_{M,Z^n}(m,z^n) - P_M(m) P_{Z^n}(z^n)
\notag \\
&\geq 
\sum_{(m,z^n) \in \mcf{M} \times \mcf{Z}^n \setminus \Omega_n(0) } 
P_{M,Z^n}(m,z^n) \left( 1 - 2^{-n \lambda_n} \right) 
\notag \\
&= 
\left[ 1 - 2^{-n \lambda_n} \right] 
\left[1-P_{M,Z^n}\left(\Omega_n(0) \right)\right].
\label{eq:sc2}
\end{align}
Thus combining Lemma~\ref{lem:apx} and~\eqref{eq:sc2} gives 
\[
P_{Q_n}\left( \mcf{Q}_n^S (0)\right) \geq 1 - \delta -
2^{-n \lambda_n} - \mu_n .
\]
As a result, if $\epsilon + \delta < 1$, then there must exist a
$q_n \in \mcf{Q}_n$ such that~\eqref{eq:wtc_1} and~\eqref{eq:wtc_2}
are simultaneously satisfied since
\[
  P_{Q_n}( \mcf{Q}_n^R \cap \mcf{Q}_n^S(0) ) \geq 1 - \epsilon -
  \delta - \zeta_n - 2^{-n \lambda_n}-\mu_n> 0
\]
for all sufficiently large $n$. And therefore, \[R \leq C(0) + \zeta_n + \tau_n\] if $\epsilon + \delta < 1$.
If though $\epsilon + \delta \geq 1$, then the strong Fano's inequality
(i.e.,~\eqref{eq:wtc_1}) gives $R \leq C(\infty) + \zeta_n$.

\subsection{Proof of Theorem~\ref{thm:3}}

\textbf{(Direct)} The RES-triple $\left((C(r_l),\epsilon, r_l \right)$ is $\mbf{S}_3$ since by definition $\left((C(r_l),0, r_l \right)$ is $\mbf{S}_3$ achievable. 

\textbf{(Converse)} On the other hand, to prove that $\mathbb{C}_{3}(r_l)$ is an upper
bound on the $\epsilon$-secrecy capacity under $\mbf{S}_3(nr_l)$ of
the DM-WTC, we note that Lemma~\ref{lem:apx} and $\mbf{S}_3(nr_l)$
directly imply
\begin{equation}
P_{Q_n}\left( \mcf{Q}_n^S(r_l) \right) \geq
P_{M,Z^n}\left(\Omega_n(r_l) \right) - \mu_n \geq 1 - \rho_n -\mu_n
\label{eq:sc3}
\end{equation}
for some $\rho_n \rightarrow 0$. Thus as before the strong Fano's
inequality and~\eqref{eq:sc3} together give the existence of a
$q_n \in \mcf{Q}_n$ satisfying~\eqref{eq:wtc_1} and
\begin{equation}
\frac{1}{n} I(M;Z^n|Q_n=q_n)  \leq  r_k + \tau_n \label{eq:wtc_3}
\end{equation}
since
$P_{Q_n}\left( \mcf{Q}_n^R \cap \mcf{Q}_n^S(r_l)\right) \geq 1 -
\epsilon - \zeta_n - \rho_n - \mu_n> 0$. Now
\[
R < C(r_l) + \zeta_n + \tau_n
\]
for all $\epsilon \in [0,1)$, follows directly as a result of Equations~\eqref{eq:wtc_1} and~\eqref{eq:wtc_3}.

\subsection{Proof of Theorem~\ref{thm:4}}

\textbf{(Direct)} First note the RES-triple $\left(C\left( \frac{r_l}{1-\epsilon} \right),0, \frac{r_l}{1-\epsilon} \right)$ is $\mbf{S}_4$ achievable due to~\cite[Theorem~17.13]{CK}. Hence the RES-triple $\left(C\left( \frac{r_l}{1-\epsilon} \right),\epsilon, r_l \right)$ is $\mbf{S}_4$-achievable by Lemma~\ref{lem:only_other_lemma}. 

\textbf{(Converse)} To prove $\mathbb{C}_{4}(\epsilon, r_l)$ upper-bounds the
$\epsilon$-secrecy capacity under $\mbf{S}_4 (nr_l)$ of the DM-WTC,
notice that $\mbf{S}_4 (nr_l)$ implies
\begin{align}
r_l 
&\geq 
\frac{1}{n}  I(M;Z^n) 
\geq \frac{1}{n}  I(M;Z^n | Q_n) - \frac{\alpha}{n} \logt n
\notag \\
&\geq
\sum_{q_n \in \mcf{Q}_n^R} \frac{1}{n} I(M;Z^n|Q_n = q_n)
  P_{Q_n}(q_n) - \frac{\alpha}{n} \logt n\notag \\
&\geq
\min_{q_n \in \mcf{Q}_n^R} \frac{1}{n} I(M;Z^n|Q_n = q_n)
  P_{Q_n}(\mcf{Q}_n^R) - \frac{\alpha}{n} \logt n
\label{eq:lk1}
\end{align}
where $n^{\alpha}$ is the cardinality bound on $\mcf{Q}_n$.
But from the strong Fano's inequality, we have
$P_{Q_n}(\mcf{Q}_n^R) \geq 1 - \epsilon - \zeta_n$. This together
with~\eqref{eq:lk1} implies that there must be a $q_n \in \mcf{Q}_n^R$
such that
\begin{equation}
\frac{1}{n} I(M;Z^n|Q_n = q_n) \leq \frac{r_l + \frac{\alpha}{n} \logt
  n}{1-\epsilon - \zeta_n}.
\label{eq:lk2}
\end{equation}
Again by the strong Fano's inequality, for this $q_n$ we also
have~\eqref{eq:wtc_1}. Combining~\eqref{eq:wtc_1}
and~\eqref{eq:lk1} gives
\[
R \leq C\left( \frac{r_l}{1-\epsilon} \right) + \frac{\zeta_n  r_l + ( 1+ \zeta_n) \frac{\alpha}{n} \logt n   }{(1- \epsilon - \zeta_n)(1-\epsilon)} + \zeta_n.
\]

\section{Conclusions}
Employing the recently developed techniques of equal-image-size
partitioning, we obtained the $\epsilon$-secrecy capacities under
$\mbf{S}_1(\delta)$, $\mbf{S}_2(\delta)$, $\mbf{S}_3(nr_l)$, and
$\mbf{S}_4(nr_l)$ of the DM-WTC for non-vanishing $\epsilon$,
$\delta$, and $r_l$. The secrecy criteria considered include the
standard leakage and variation distance secrecy constraints often
employed in the literature. Our new results show that both the
capacity value and the strong converse property of the DM-WTC are in
fact dependent on the secrecy criterion adopted. We conjecture that
the interesting phase change phenomenon observed in cases where the
strong converse property does not hold is commonplace in many other
multi-terminal DMCs.

\appendix

\subsection{Proof of Lemma~\ref{lem:apx}}
We need the following lemma to prove Lemma~\ref{lem:apx}: 
\begin{lemma} \label{lem:only_lemma} Let $Q_n$ be a random index
  ranging over $\mcf{Q}_n$, whose cardinality is at most polynomial in
  $n$, and $V$ be any discrete random variable distributed over
  $\mcf{V}$. Then there exist $\lambda_n \rightarrow 0$ and
  $\xi'_n \rightarrow 0$ such that $n\lambda_n \rightarrow \infty$ and
\begin{align*}
& \hspace*{-5pt}
P_{V,Q_n} \left( \set{ (v,q_n) \in \mcf{V}\times \mcf{Q}_n :
  P_{V|Q_n}(v|q_n) \eo{\lambda_n} P_V(v)}  \right) 
\notag \\
&\geq 1-\xi'_n.
\end{align*}
Note that $\lambda_n$ and $\xi'_n$ both depend only on the polynomial
cardinality bound on $\mcf{Q}_n$.
\end{lemma}
\begin{IEEEproof}
Let $\alpha > 0$ be such that $\abs{\mcf{Q}_n} \leq n^{\alpha}$. 
First write $\mcf{A} = \set{(v,q_n) \in  \mcf{V}\times \mcf{Q}_n:
  P_{V|Q_n}(v|q_n) > n^{2\alpha} P_V(v) }$ and 
$\mcf{B} = \set{(v,q_n) \in  \mcf{V}\times \mcf{Q}_n:
  P_{V|Q_n}(v|q_n) < n^{-2\alpha} P_V(v)}$. Then 
\begin{align}
& \hspace*{-5pt} 
P_{V,Q_n} \left( \set{ (v,q_n) \in \mcf{V}\times \mcf{Q}_n :
  P_{V|Q_n}(v|q_n) \eo{\lambda_n} P_V(v)}  \right) 
\notag \\
&\geq 1 - P_{V,Q_n}\left( \mcf{A} \cup \mcf{B} \right)
\label{eq:apx1}
\end{align}
where $\lambda_n = \frac{2\alpha}{n} \logt n$. Thus the lemma is
verified by~\eqref{eq:apx1} if we can show that
$P_{V,Q_n}\left( \mcf{A} \cup \mcf{B} \right) \rightarrow 0$. In
particular, we do so by bounding $P_{V,Q_n} (\mcf{A}) \leq n^{-\alpha}$
and $P_{V,Q_n} (\mcf{B}) \leq n^{-2\alpha}$, and setting
$\xi'_n = n^{-\alpha} + n^{-2\alpha}$.

To bound $P_{V,Q_n} (\mcf{A})$, note that for all  $(v,q_n) \in
\mcf{A}$, 
\begin{equation}
P_{Q_n}(q_n) \leq n^{-2\alpha} 
\label{eq:ql1}
\end{equation}
since
\[
P_V(v) \geq P_{V|Q_n}(v|q_n) P_{Q_n}(q_n) \geq n^{2\alpha}P_V(v)
P_{Q_n}(q_n). 
\]
Then the upper bound on $P_{V,Q_n} (\mcf{A})$ follows
from~\eqref{eq:ql1} as below:
\begin{align*}
P_{V,Q_n} (\mcf{A}) 
&= \sum_{(v,q_n) \in \mcf{A} } P_{V|Q_n}(v|q_n) P_{Q_n}(q_n)
\\
&\leq \sum_{(v,q_n) \in \mcf{A} } P_{V|Q_n}(v|q_n) n^{-2\alpha} 
\leq n^{-\alpha}.
\end{align*}
The upper bound on $P_{V,Q_n} (\mcf{B})$ follows similarly in that
\begin{align*}
P_{V,Q_n} (\mcf{B})
&= \sum_{(v,q_n) \in \mcf{B} } P_{V|Q_n} (v|q_n) P_{Q_n}(q_n)
\\
&\leq \sum_{(v,q_n) \in \mcf{B} } P_{V} (v) P_{Q_n}(q_n) n^{-2\alpha} 
\leq n^{-2\alpha}.
\end{align*}
\end{IEEEproof}

Apply Lemma~\ref{lem:only_lemma} three times with $V=M$, $V=Z^n$, and $V =
(M,Z^n)$, respectively. Writing 
\begin{align*}
\Gamma_n
 &\defn 
 \Big\{ (m,z^n,q_n) \in \mcf{M}\times\mcf{Z}^n\times\mcf{Q}_n:
\\
&\hspace*{30pt}
P_{M,Z^n|Q_n}(m,z^n|q_n) \eo{\lambda_n} P_{M,Z^n}(m,z^n),
\\
&\hspace*{30pt}
P_{M|Q_n}(m|q_n) \eo{\lambda_n} P_{M}(m), \text{ and}
\\
&\hspace*{30pt}
P_{Z^n|Q_n}(z^n|q_n) \eo{\lambda_n} P_{Z^n}(z^n) \Big\}
\end{align*}
where $\lambda_n$ is obtained in Lemma~\ref{lem:only_lemma}, we have 
\begin{equation}\label{eq:almost|Q}
  P_{M,Z^n,Q_n}(\Gamma_n) \geq 1 - 3\xi'_n.
\end{equation}
Next define
\begin{align*}
\Xi_n 
 &\defn 
\Big\{(m,z^n,q_n) \in \mcf{M}\times\mcf{Z}^n\times\mcf{Q}_n :
   q_n \in \mcf{Q}^Z_n, 
\\
& \hspace{40pt}
  m \in \mcf{\tilde M}(q_n), \text{ and } z^n \in \mcf{\hat Z}^n(q_n)
  \cap \mcf{\tilde Z}^n(m,q_n)
 \Big\}
\end{align*}
with the corresponding $\mcf{Q}^Z_n$, $\mcf{\tilde M}(q_n)$,
$\mcf{\hat Z}^n(q_n)$, and $\mcf{\tilde Z}^n(m,q_n)$ as given in the
information stabilization result summarized in
Section~\ref{sec:proofs}. Similar to before, 
\begin{equation}\label{eq:almostU}
  P_{M,Z^n,Q_n}(\Xi_n) \geq 1 - 4\xi_n.
\end{equation}
Combining~\eqref{eq:almost|Q} and~\eqref{eq:almostU} gives
\begin{equation}\label{eq:almost}
  P_{M,Z^n,Q_n}(\Xi_n \cap \Gamma_n) \geq 1 - 3\xi'_n  - 4\xi_n.
\end{equation}
From here note that for any $(m,z^n,q_n) \in \Xi_n \cap \Gamma_n$,
\[
P_{M,Z^n}(m,z^n)\leq 2^{n (r+\lambda_n)} P_{M}(m) P_{Z^n}(z^n)
\]
implies
\begin{align}
\cexp{\frac{P_{Z^n,M|Q_n}(z^n,m|q_n)}{P_{Z^n|Q_n}(z^n|q_n)P_{M|Q_n}(m|q_n) }}  &\leq r+ 4 \lambda_n,
\label{eq:apx3}
\end{align}
because $(m,z^n,q_n) \in \Gamma_n$. And then in turn, for all $(m,z^n,q_n) \in \Gamma_n \cap \Xi_n$, 
\begin{align}
r+4 \lambda_n 
&\geq
\frac{1}{n} I(M;Z^n|Q_n=q_n) - 2\xi_n
\label{eq:apx4}
\end{align}
since $(m,z^n,q_n) \in \Xi_n$.
Thus Lemma~\ref{lem:apx} results from~\eqref{eq:apx4} by setting
$\tau_n = 4 \lambda_n + 2\xi_n$ and $\mu_n  = 3\xi'_n  + 4\xi_n$, 
because we have from~\eqref{eq:almost}
\begin{align*}
& \hspace*{-10pt}
P_{M,Z^n}(\Omega_n (r) )
\\
&\leq
P_{M,Z^n,Q_n}\left( \Xi_n \cap \Gamma_n \cap \Omega_n (r)
  \times \mcf{Q}_n \right) + 3\xi'_n  + 4\xi_n 
\\
& \leq P_{Q_n} \left(\mcf{Q}_n^S(r) \right) +
  3\xi'_n  + 4\xi_n.
\end{align*}

\subsection{Proof of Lemma~\ref{lem:only_other_lemma}}

For $i \in \set{2,4}$, we can construct a $\left(n,R_n,(1-\gamma)\epsilon_n + \gamma, \mbf{S}_i \left( (1- \gamma)l_n \right)\right)$-code $(\hat f^n,\varphi^n)$, given that there exists a $(n,R_n,\epsilon_n,\mbf{S}_i(l_n) )$-code $(f^n,\varphi^n)$. Whence the lemma follows by the definition of the RES-triples. Letting
$\hat M$ be a random variable distributed identical, but independent,
to $M$. The new encoder, $\hat f^n$, is constructed by setting it equal to $f(M)$ with probability $1-\gamma$ and to $f(\hat M)$ with probability $\gamma$. While the new decoder $\hat \varphi^n = \varphi^n$. 

Clearly, an error will likely occur if $\hat f(M)$ is set equal to $f(\hat M)$. On the other hand, the probability of error will revert to that of $(f^n,\varphi^n)$ if $\hat f(M)$ is set equal to $f(M)$. Thus the probability of error for $(\hat f^n,\hat \varphi^n)$ is at most
$(1-\gamma)\epsilon_n + \gamma$. 

Letting $P_{Z^n,M}$ be the joint
distribution of $Z^n,M$ for induced by $f^n$, we can write the
joint distribution of $Z^n,M$ for $\hat f^n$ as
$(1-\gamma) P_{Z^n,M} + \gamma P_{Z^n}P_{M}$, while the marginals
remain $P_{M}$ and $P_{Z^n}$. But then, for the variation distance,
\begin{align*}
&\labs{(1-\gamma) P_{Z^n,M} + \gamma P_{Z^n}P_{M} - P_{Z^n}P_{M} }\\
&\hspace{20pt}=(1-\gamma) \labs{ P_{Z^n,M}  - P_{Z^n}P_{M} }.
\end{align*}
And for divergence 
\begin{align*}
&D\left( (1-\gamma) P_{Z^n,M} + \gamma P_{Z^n}P_{M} \middle| \middle| P_{Z^n}P_{M} \right)\\
&\hspace{10pt}\leq (1-\gamma) D\left( P_{Z^n,M} \middle| \middle| P_{Z^n}P_{M} \right) + \gamma D\left(  P_{Z^n}P_{M} \middle| \middle| P_{Z^n}P_{M} \right) \\
&\hspace{10pt}= (1-\gamma) D\left( P_{Z^n,M} \middle| \middle| P_{Z^n}P_{M} \right) .
\end{align*}

\bibliographystyle{ieeetr} 

\begin{thebibliography}{1}

\bibitem{Wyner1975}
A.~Wyner, ``The wire-tap channel,'' {\em Bell Syst. Tech. J.}, vol.~54,
  pp.~1355--1387, Oct. 1975.

\bibitem{bloch13strong}
M.~Bloch and J.~Laneman, ``Strong secrecy from channel resolvability,'' {\em
  IEEE Trans. Info. Theory}, vol.~59, pp.~8077--8098, Dec. 2013.

\bibitem{tan14wiretap}
V.~Tan and M.~Bloch, ``Information spectrum approach to strong converse
  theorems for degraded wiretap channels,'' in {\em Proc. 52nd Annual Allerton
  Conf. Comm., Con., and Comp.}, pp.~747--754, Sep. 2014.

\bibitem{graves2015equal}
E.~Graves and T.~F. Wong, ``Equal-image-size source partitioning: Creating
  strong fano's inequalities for multi-terminal discrete memoryless channels,''
  {\em ArXiv e-prints}, Dec. 2015.
\newblock Available at \texttt{https://arxiv.org/abs/1512.00824 }.

\bibitem{2016arXiv161004215W}
Y.-P. {Wei} and S.~{Ulukus}, ``{Partial Strong Converse for the Non-Degraded
  Wiretap Channel},'' {\em ArXiv e-prints}, Oct. 2016.
\newblock Available at \texttt{https://arxiv.org/abs/1610.04215 }.

\bibitem{Hayashi14}
M.~Hayashi, H.~Tyagi, and S.~Watanabe, ``Strong converse for a degraded wiretap
  channel via active hypothesis testing,'' in {\em Proc. 52nd Annual Allerton
  Conf. Comm., Con., and Comp.}, pp.~148--151, Sep. 2014.

\bibitem{graves2016information}
E.~Graves and T.~F. Wong, ``Information stabilization of images over discrete
  memoryless channels,'' in {\em Proc. IEEE Int. Symp. Info. Theory},
  pp.~2619--2623, Jun. 2016.

\bibitem{CK}
I.~Csisz{\'a}r and J.~K{\"o}rner, {\em Information Theory: Coding Theorems for
  Discrete Memoryless Systems}.
\newblock Cambridge University Press, 2nd~ed., 2011.

\end{thebibliography}

\end{document}